\documentclass[conference,letterpaper]{IEEEtran}
\IEEEoverridecommandlockouts
\usepackage{cite}
\usepackage{amsmath,amssymb,amsfonts}
\usepackage{algorithmic}
\usepackage{graphicx}
\usepackage{textcomp}
\usepackage{xcolor}
\usepackage{hyperref}
\usepackage{caption}
\usepackage{enumitem}

\usepackage{todonotes}
\usepackage{pdfsync}

\usepackage{lipsum}
\usepackage{colortbl}
\newcolumntype{G}{>{\columncolor[gray]{0.8}[.5\tabcolsep][\tabcolsep]}c}
\usepackage{epstopdf}
\usepackage{listings}
\newcommand{\muop}{\mbox{$\mu$-op}}
\newcommand{\muops}{\mbox{$\mu$-ops}}

\newcommand{\highl}[2]{\textcolor{#1}{\textbf{#2}}}

\def\BibTeX{{\rm B\kern-.05em{\sc i\kern-.025em b}\kern-.08em
    T\kern-.1667em\lower.7ex\hbox{E}\kern-.125emX}}
\begin{document}

\title{{Automated Instruction Stream Throughput Prediction for Intel and AMD Microarchitectures}
\thanks{This work was partly funded by BMBF through the METACCA project.}}

\author{\IEEEauthorblockN{
Jan Laukemann\IEEEauthorrefmark{1},
Julian Hammer\IEEEauthorrefmark{2},
Johannes Hofmann\IEEEauthorrefmark{3},
Georg Hager\IEEEauthorrefmark{4} and Gerhard Wellein\IEEEauthorrefmark{5}}
\IEEEauthorblockA{Friedrich-Alexander-Universit\"at Erlangen-N\"urnberg\\
Erlangen, Germany}
\IEEEauthorblockA{\IEEEauthorrefmark{1}\textit{Department of Computer Science}\\
jan.laukemann@fau.de}
\IEEEauthorblockA{\IEEEauthorrefmark{2}\textit{Erlangen Regional Computing Center}\\
julian.hammer@fau.de}
\IEEEauthorblockA{\IEEEauthorrefmark{3}\textit{Chair of Computer Architecture} \\
johannes.hofmann@fau.de}
\IEEEauthorblockA{\IEEEauthorrefmark{4}\textit{Erlangen Regional Computing Center} \\
georg.hager@fau.de}
\IEEEauthorblockA{\IEEEauthorrefmark{5}\textit{Erlangen Regional Computing Center} \\
gerhard.wellein@fau.de}}

\maketitle

\begin{abstract}
    
An accurate prediction of scheduling and execution of instruction
streams is a necessary prerequisite for predicting the in-core
performance behavior of throughput-bound loop kernels on out-of-order
processor architectures. Such predictions are an indispensable
component of analytical performance models, such as the Roof\/line and
the Execution-Cache-Memory (ECM) model, and allow a deep understanding
of the performance-relevant interactions between hardware architecture
and loop code.

We present the Open Source Architecture Code Analyzer (OSACA),
a static analysis tool for predicting the execution time of
sequential loops comprising x86 instructions under the assumption of
an infinite first-level cache and perfect out-of-order scheduling. We
show the process of building a machine model from available documentation and
semi-automatic benchmarking, and carry it out for the latest Intel~Skylake and
AMD~Zen micro-architectures.

To validate the constructed models, we apply them to several assembly
kernels and compare runtime predictions with actual
measurements. Finally we give an outlook on how the method may be
generalized to new architectures.

\end{abstract}


\section{Introduction}\label{sec:intro}

Looking at numerical codes, compute-intensive applications and the resources (time, energy, hardware)
they consume, it is vital to optimize them for performance in order to reduce their resource
consumption. One of the most fundamental ways of approaching this is performance modeling, where a
(simplified) model of the underlying hardware is used to predict the runtime of a
computational kernel. The Roof\/line~\cite{Roofline}
and ECM~\cite{ECM} performance models are probably the most common tools
employed for this task on modern CPUs.
When applying them, a performance-aware developer will start to build an
understanding of the characteristics of the architecture-code interactions, and the model will
pinpoint the constraining bottleneck. Once known, the bottleneck can often be mitigated
by changes in the code, the runtime parameters, or the execution environment.
When the models' construction is automated~\cite{RooflineModelToolkit,Kerncraft}, compilers and
a wider user base can take advantage of them.

In practice, the analysis and modeling process on a given architecture
is typically split in two parts: in-core execution and data
transfer. For example, in the simplest form of the Roof\/line model,
calculating the chip's maximum performance taking only the floating-point
operations into account is the in-core execution analysis while
deriving the arithmetic intensity relies on data transfer analysis. In
this work we focus on a refined in-core execution analysis, where the
essential questions is: How many cycles does it take \emph{at least} to execute a set
of assembly instructions that constitute the body of an infinite
loop? The resulting cycle count yields an absolute upper
performance bound (or roof), and it is valid for all processor models of
the same microarchitecture. This is not the same as counting FLOPs,
but a similar and more realistic approach, which may also be applied
to non-floating-point codes~\cite{Kerncraft}.

The in-core analysis makes a number of assumptions, which will be
explained later in further detail.
Intel already provides the Intel Architecture Code Analyzer
(IACA)~\cite{iaca}, an in-core static analyzer for their latest
architectures. It has proven extremely valuable for analytic performance
modeling. Unfortunately, IACA is both closed-source and restricted to
Intel CPUs. We want to develop an open version,
with which developers can not only see the analysis outcome but also the
underlying model. Beyond the tool itself we want to extend our approach
to other, non-Intel architectures and platforms.

This paper is organized as follows: In Sections~\ref{ssec:background} through
\ref{ssec:validation_hardware}, we elaborate on the assumptions stated above, give a general
overview of the hardware model and describe relevant features of our example architectures and the
hard\-ware/soft\-ware environment. Section~\ref{ssec:related} covers related work. In
Section~\ref{sec:methodology} we explain how to build a detailed machine model for an architecture
from available documentation and benchmarking. We exercise this methodology on our example
architectures in Section~\ref{ssec:methodology_example}. Section~\ref{sec:implementation} explains
technical details about the static analyzer and compare its predictions with actual measurements in
Sections~\ref{ssec:implementation_example} and \ref{ssec:implementation_example_2}. Finally,
Section~\ref{sec:conclusion} summarizes the work and gives an outlook to future developments.

The OSACA software is available for download at \cite{osaca}. Information about how to reproduce
the results in this paper can be found in the artifact description~\cite{artif}.

\subsection{Background}\label{ssec:background}

\begin{figure}[tbp]
\centerline{\includegraphics[width=0.6\columnwidth]{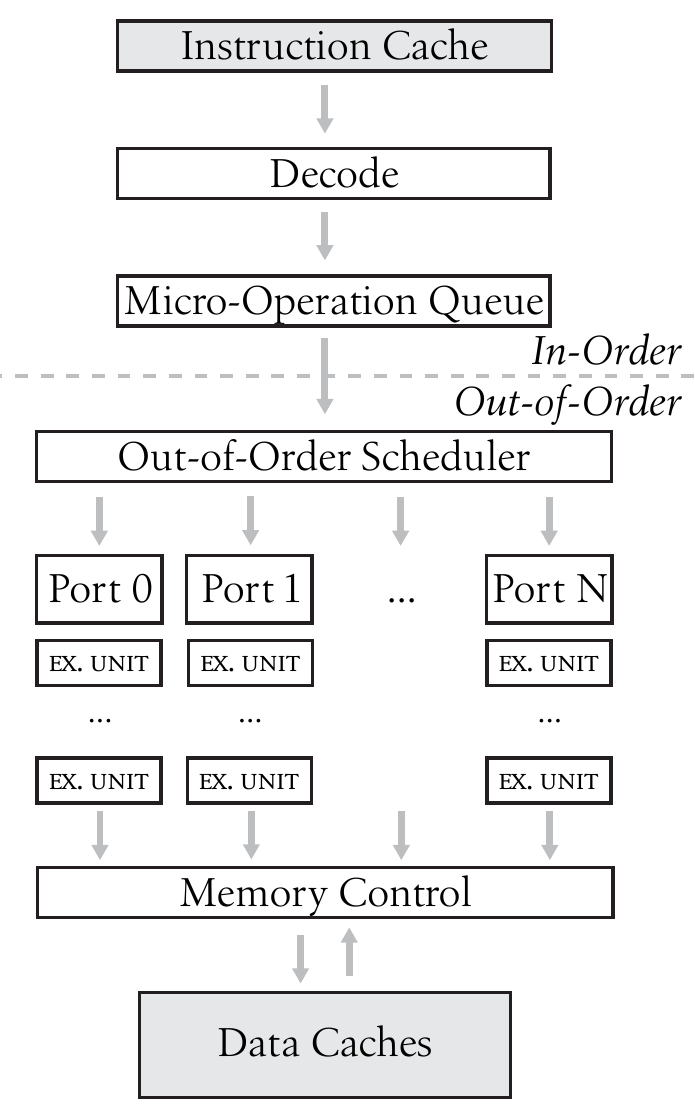}}
\caption{Assumed generic out-of-order port model for modeling, benchmarking and analysis. Functional units (e.g., ALU, AGU, MUL, DIV) are associated with ports. An out-of-order scheduler assigns \muops\ to ports, which then use their functional units to execute the instructions in an pipelined fashion.}
\label{fig:generic_portmodel}
\end{figure}

When thinking about the performance of a CPU core, we assume what is widely known as the ``port
model'': each instruction is (optionally) split into \emph{micro-ops} (\muops), which get executed
by functional units. A particular instruction may have multiple functional units that can execute it
(e.g., two integer ALUs), or -- in case of complex instructions -- multiple functional units
\emph{must} execute it (e.g., combined load and floating-point add). Functional units are grouped
behind ports, with one port serving one or more units. Each port can receive only one instruction
per cycle. Figure~\ref{fig:generic_portmodel} shows a diagram of such a generic port model.

The following assumptions, already stated in Section~\ref{sec:intro}, are assumed for our
prediction model:
\newline
\begin{enumerate}
    \item \emph{All data accesses hit the first-level cache.}
    
          This is where the boundary between in-core and data analysis is drawn. If a dataset fits
          in the first-level cache, all accesses will behave the same and there is no need to
          consider the order and pattern of previous accesses or (possibly undisclosed) cache
          replacement algorithms. Behavior beyond L1 can be modeled separately, but this is beyond
          the scope of this work (the Kerncraft tool~\cite{Kerncraft}, which relies on an in-core
          analysis from IACA and -- in the future -- OSACA, combines it with data analysis for
          a unified Roof\/line or ECM prediction).
        
    \item \emph{Multiple available ports per instruction are utilized with fixed probabilities.}
        
          Since the actual scheduling algorithm is unknown, we assume that all suitable ports for
          the same instruction are used with fixed probabilities. E.g., an add instruction that may
          use one of two ports may be scheduled half the time on one and half the time on the
          other, or one-tenth of the time on one and nine-tenth of the time on another.
          Consideration of actual port pressure is currently not supported yet, but may be
          considered when it becomes necessary to better mimic measured performance.
    
    \item \emph{Otherwise, out-of-order scheduling by the hardware works perfectly.}
    
          The previous assumption implies imperfect scheduling if ports are asymmetric. Asymmetry
          means that multiple ports can handle the same instruction, but other features of those
          ports differ (e.g., one port supports \texttt{add} and \texttt{div}, while another
          supports \texttt{add} and \texttt{mul}). This may cause load imbalance since, e.g., a
          code with only \texttt{add} and \texttt{mul} may be imperfectly scheduled. Since the
          actual scheduling scheme is unknown and can only be inferred by thorough measurements,
          reverse engineering the details of the scheduling algorithms
          is left for future work.
          
    \item \emph{All latencies are hidden via speculative execution.}
          
          Speculative execution and out-of-order scheduling allows the processor to execute a loop
          kernel with intra-iteration dependencies as a throughput-bound code (i.e., the
          pipeline which is the bottleneck is fully utilized). In other words, the critical
          execution path through the loop iteration can be ignored. Similar to IACA, we focus on
          throughput modeling at the moment and do not model latency.
          
\end{enumerate}

To the best of our knowledge, assumptions 1, 3 and 4 apply to IACA
as well, but due to the undisclosed machine model behind IACA we are unable to validate this. For
assumption 2, IACA shifts probabilities to balance port pressures.
In Section~\ref{sec:methodology}, we will go into detail
about how we derive our model parameters from available sources and benchmarking.

Available, but incomplete and sometimes misleading sources are: architecture diagrams and
performance numbers found in technical manuals~\cite{IntelOptMan} and marketing
presentations~\cite{AMDMarketingFoo} of vendors, third-party researchers \cite{diamond11} and
enthusiasts~\cite{Agner} compiling their own benchmarking results.

\subsection{Intel~Skylake and AMD~Zen Architectures}\label{ssec:architectures}

Comprehensive information is available on Intel's micro-architectures, and we therefore have a clear
understanding of the overall behavior. We will now go into performance-relevant details on Intel
Skylake, followed by a discussion of AMD~Zen.

On Intel~Skylake, each port ($0-7$) can consume one \muop\ per cycle. A \muop\ may take
any number of cycles to retire. Simple instructions (e.g., \texttt{vaddpd \%xmm1,\%xmm2,\%xmm3} or
``add values in \texttt{xmm1} and \texttt{xmm2} and store result to \texttt{xmm3}'') map to
exactly one \muop, while complex instructions are split into multiple \muops\ 
(e.g., \texttt{vaddpd \%xmm1,(\%eax),\%xmm3} or ``load values at memory address \texttt{eax}, add
with values in \texttt{xmm1} and store result to \texttt{xmm3}'').\footnote{Unless otherwise noted,
  we use the AT\&T (destination last) form of the x86 assembly syntax here. IACA uses Intel syntax
  (destination first) in its output.}

\begin{figure}[tbp]
\centerline{\includegraphics[width=\columnwidth]{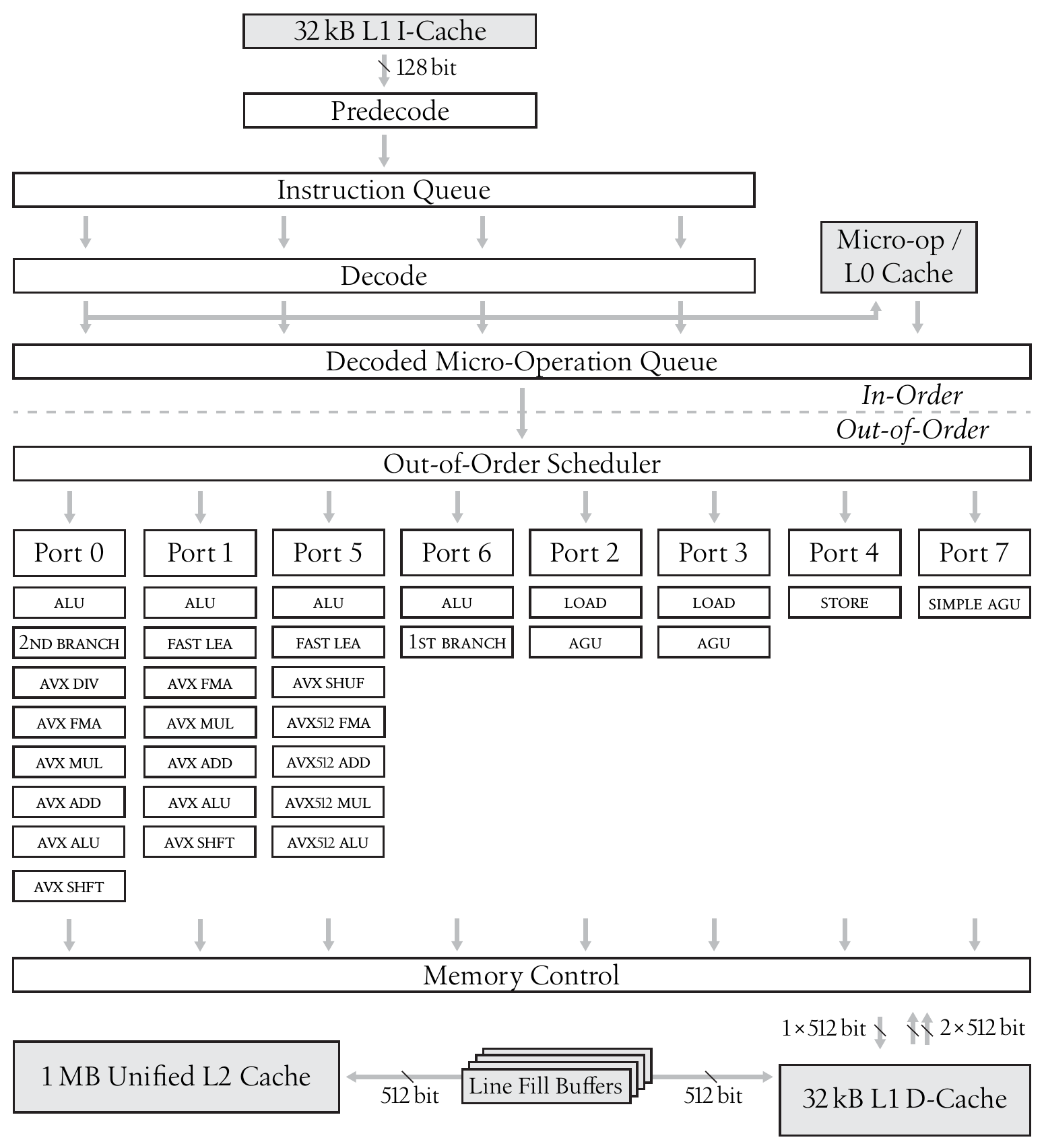}}
\caption{Intel~Skylake core block diagram and 
  port model, compiled from Intel's Optimization Manual~\cite{IntelOptMan}.}
\label{fig:skl_diagram}
\end{figure}

In Figure~\ref{fig:skl_diagram} we see the mapping of ports to functional units and thus
instructions. Scalar integer instructions need either port 0, 1, 5 or 6. $256\,\mathrm{bit}$ wide
vector instructions go to port 0 or 1. Divides are always handled by port 0.
Loads occupy port 2 or 3, and stores need port 4 as well as 2, 3 or 7 for address calculations.

In addition to ports, there are other potential bottlenecks, in particular instruction cache
bandwidth and fetch and decode throughput:  The L1
instruction cache is limited to $32\,\mathrm{KiB}$ and can serve $16\,\mathrm{Bytes}$ per cycle to
the fetcher. The decoders can emit a total of five \muops\ per cycle, four from simple
instructions and one from a complex instruction. Currently we ignore those limits.

During allocation and renaming, architectural register IDs from the machine code are
replaced with physical registers. In
combination with move elimination and zeroing idioms (also during the allocation and renaming
step), the processor is able to locate and circumvent false data dependencies. All
independent instructions can then be scheduled on ports providing the necessary functional units.

One new instruction can be scheduled on each port per cycle; however, some special conditions
exist. One prominent example, which we also model, is: Divide instructions are executed scheduled
on port 0, and they take four cycles, but the port already becomes available to non-divide
instructions on the next cycle. We, as well as IACA, model this using an additional port called
0DV, which only handles divide instructions and is occupied for four cycles, while port 0 is only
occupied for 1 cycle.

Loads go through ports 2 and 3. Both ports also lead to the necessary address generation units
(AGUs). The ``store port'' (4) does not come with its own AGU, thus each store requires an AGU from
port 2 or 3, or -- if the address is simple -- from port 7.
\newline
\newline
\begin{figure}[tbp]
\centerline{
\includegraphics[width=\columnwidth]{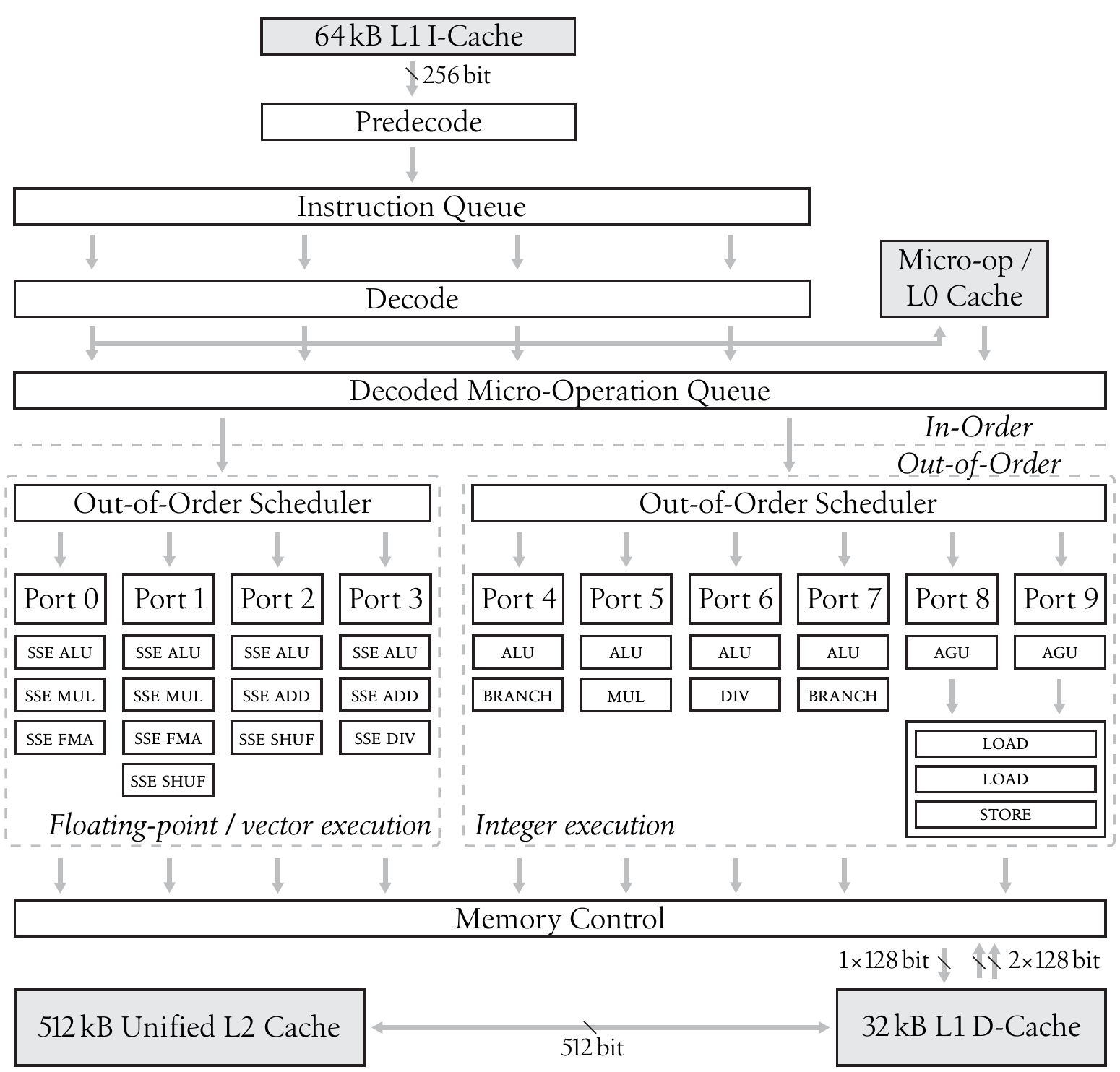}}
\caption{AMD Zen core block diagram and port model, compiled from AMD's Optimization
Guide~\cite{AMDOptMan}, marketingslides~\cite{AMDMarketingFoo} and Agner Fog's Instruction
Tables~\cite{Agner}.}
\label{fig:zen_diagram}
\end{figure}

\subsection{Validation Hardware, Software and Runtime Environment}\label{ssec:validation_hardware}

All results presented were gathered on two machines:
\begin{description}[labelwidth=1.3cm,leftmargin=1.5cm]
    \item[Skylake] Intel Xeon i7-6700HQ with Skylake micro-architecture running at fixed 1.8\,GHz with turbo disabled
    \item[Zen] AMD EPYC 7451 with Zen micro-architecture, running at fixed 1.8\,GHz with turbo disabled
\end{description}

OSACA (version 0.2.0) was run with Python~v3.5.3 and benchmarks were compiled using GCC
7.2.0. When compiling for Intel
Skylake we used the flags \texttt{-fopenmp-simd -march=broadwell}. Although AVX-512 could be
modeled, we deliberately ignored this capability, since we wanted to compare prediction and
execution of the same assembly code on both architectures and AVX-512 instructions are not
supported on AMD Zen. Compiling for AMD Zen was done with \texttt{-fopenmp-simd -march=znver1
-mavx2 -mfma} compiler options. For both platforms we created different versions of the code by
using the \texttt{-O1}, \texttt{-O2} and \texttt{-O3} flags, respectively.

During execution, we used \texttt{likwid-pin} to pin the processes to a physical core. That and
fixing the frequency reduced fluctuations during runtime measurements. Leaving turbo mode enabled
would lead to unusual results, because the CPU frequency changes during execution and calculation
of cycles from a combined runtime becomes impossible. In effect, statistical runtime variations
were small enough to be ignored. In all measurements we nevertheless report the ``best'' value
(highest performance, lowest runtime).

\subsection{Related Work}\label{ssec:related}

In general, there are two approaches to predicting runtime and performance behavior: static analysis
and simulation. Our work is set in the static analysis category, because we expect results to be
explanatory in order to guide developers and tools in optimizing performance, and to be available
fast in order to allow inclusion in other tools, such as compilers. Simulators on the other hand may
be more thorough and accurate \emph{if} comprehensive implementations exist. They can also consider
the data side, such as diverging branches or interaction of multiple cores or nodes. These
advantages come at a cost: Steady states for throughput analysis need to be found, valid and
representative data needs to be available, pinpointing a bottleneck becomes non-trivial and
implementation is much more complex.

Being an inspiration for this work, the most prominent example for static analysis tools is IACA
itself~\cite{iaca}. Developed by Israel Hirsh and Gideon S. [sic], Intel released the tool in 2012
and has issued the latest version in 2017. It is closed source and the underlying model neither
been published by the authors, nor peer reviewed. The latest version supports throughput analysis
on Intel micro-architectures Haswell through Skylake (including AVX-512). It has built-in insight
on decomposition of instructions into \muops, \muop\ fusion and the port assignments. It also seems
to use a heuristic for scheduling instructions to ports, which we have no knowledge of. The
underlying model is bound to be more accurate than anything OSACA can hope for, due to undisclosed
information available to the developers and the complete focus on recent Intel architectures.
OSACA, on the other hand, can model non-Intel architectures and gives the user information about
the underlying model.

Two new projects came up recently in the LLVM community: LLVM-MCA~\cite{llvm_mca} and
LLVM-Exegesis~\cite{llvm_exegesis}. Both of them aim at enhancing and using available out-of-order
performance information in LLVM to improve instruction selection during compile time and to support
developers. LLVM-Exegesis benchmarks operations and derives latency and port assignment of
solitary instructions (i.e., not of assembly basic blocks) through hardware event counting. The
gathered information is meant to validate LLVM's TableDef scheduling models. LLVM-MCA is a
simulator that uses the available scheduling information from the backend to predict the expected
throughput of a basic block, similar to what IACA and OSACA do. Unlike the latter two, LLVM-MCA
actually runs a simulation of instructions through LLVM's backend.

Mendis et al.~\cite{Ithemal} apply a black-box machine learning approach to throughput estimation,
while also trying to capture memory hierarchy behavior beyond the first-level cache. The outcome of
their prediction is a single-number throughput estimation based on a generic deep neural network.
This is helpful to compilers when comparing possible code transformations, but is not sufficient from a
performance engineering perspective, where we are interested in the origin of the bottleneck and
hints on how to avoid it. It is also very important to us to separate the memory hierarchy from
execution effects in order to support performance modeling using the Roof\/line and ECM models. Ithemal,
their software, and the trained neural network were not publicly available at the time of writing.

Another simulator covering instruction execution is gem5~\cite{gem5}, developed by Binkert et al.
It supports many instruction set architectures (x86, ARM, Power and SPARC, among others), including
a complete memory system, multi-core, cache coherency, DMA, PCI, networks and more. It is
considered a ``full-system'' simulator, which goes above and beyond what the scope of this work is,
but is is rooted in the simulation domain. Gem5 also lacks support for important ISA extensions,
such as AVX.

ZSim by Sanches et al.~\cite{zsim} and MARSSx86 by Patel et al.~\cite{MARSSx86} are also
full-system simulators which give a coarse overview on complete systems (with thousands of cores or
machines), rather than detailed insights pinpointing at a bottleneck.

Charif-Rubial et al. introduced CQA~\cite{cqa}, a performance static analysis tool focused on
single-core performance of loop-centric code. It is not their goal to predict runtime, but rather
give the developer a quality estimate of the code based on static binary analysis. While they also
use benchmarks to determine instruction throughput and latency, they have opted for not modeling
out-of-order execution.

\section{Model-Construction Methodology}\label{sec:methodology}

To construct a suitable port model for a given CPU architecture, we need to identify the relevant
ports for throughput and latency during execution, as well as any other functional units occupied.
Additional non-bottleneck units do not influence the runtime of an instruction (the latency is
hidden by the bottleneck), but they may become a bottleneck when used in combination with other
instructions simultaneously. Identification of hidden non-bottleneck ports can be achieved by
combined benchmarking of multiple instructions. In the following sections, we will explain this
approach in detail for the latest AMD~Zen and Intel~Skylake architectures. Further on, we show how
to integrate the gained knowledge into OSACA's database \cite{osaca} for a throughput prediction
model.

Since the definition of ``instruction'' is ambiguous, we introduce the term \emph{instruction
form}~\cite{jansba}, which refers to an assembly instruction together with their operand types.
E.g., \texttt{vaddpd} may be used with 128\,bit, 256\,bit or 512\,bit registers, and memory
operands and an optional masking register. The types of operands have an impact on the resulting
performance and therefore need to be considered. \texttt{vaddpd mem,xmm,xmm} is the instruction
form of \texttt{vaddpd} with a source memory reference, a 128\,bit source register
and a 128\,bit target register.

Although OSACA is capable of distinguishing between different ways of addressing the memory (detecting
base, offset, index, scale factor and segment registers), in the current stage of development a
separation regarding the benchmark measurement and therefore the port distribution is not provided.
Hence, we assume that the maximum throughput of an instruction is independent of its memory addressing
mode.

\subsection{Benchmarking Latency and Throughput} \label{ssec:latTp}

To obtain the latency and throughput of an instruction, we automatically create assembly benchmarks
for use with ibench~\cite{ibench}. It offers the infrastructure to initialize, run and accurately measure the
desired parameters.

For latency benchmarking we create a dependency chain by
using the destination register of one
instruction as a source register for the next and embedding a suitable
number of back-to-back instructions into a loop.
A benchmark code for the latency of \texttt{vaddpd} may look as follows:
\begin{lstlisting}[frame=lines]
loop:
  inc       %eax 
  vaddpd    %xmm0, %xmm1, %xmm0
  vaddpd    %xmm1, %xmm0, %xmm0
  vaddpd    %xmm0, %xmm1, %xmm0
  ...
  vaddpd    %xmm1, %xmm0, %xmm0
  cmp %eax, %edx   # loop count
  jl loop
\end{lstlisting}
The above code yields a latency of 4\,cy on Intel~Skylake and 3\,cycles on AMD~Zen.

For throughput measurement, instructions with independent source and destination
operands must be issued. This could be achieved by not reusing any destination registers, but
will easily exhaust all available registers. Since we do not want to rely on the register renaming
capabilities of the core to compensate for that, multiple independent dependency
chains are created to ensure that enough independent instructions are available to utilize
all functional units. The inner loop body is long enough to compensate loop overheads:
\begin{lstlisting}[frame=lines]
loop:
  inc       %eax 
  vaddpd    %xmm3, %xmm0, %xmm0
  vaddpd    %xmm4, %xmm1, %xmm1
  vaddpd    %xmm5, %xmm2, %xmm2
  vaddpd    %xmm3, %xmm0, %xmm0
  vaddpd    %xmm4, %xmm1, %xmm1
  vaddpd    %xmm5, %xmm2, %xmm2
  vaddpd    %xmm3, %xmm0, %xmm0
  vaddpd    %xmm4, %xmm1, %xmm1
  vaddpd    %xmm5, %xmm2, %xmm2
  vaddpd    %xmm3, %xmm0, %xmm0
  ...
  cmp %eax, %edx   # loop count
  jl loop  
\end{lstlisting}
This benchmark yields a throughput of 2 instructions per cycle on
Intel~Skylake and AMD~Zen. From this we can infer that
two independent ports (and thus pipelines) are available for \texttt{vaddpd xmm,xmm,xmm}.

\subsection{Benchmarking Port Occupation}


The port model, in which each port may feed multiple execution units, creates a peculiar bottleneck
when a code comprises a mixture of different instruction forms that must go through the same
port. Even though ample execution resources are available, the performance may be impeded by the
limit of one instruction per cycle and port.  This ``port conflict'' can be measured: By adding
another instruction form into the already throughput bound benchmark, either an increase or no
change in runtime is expected. If the runtime increased, both instruction forms utilize at least one
common port, which needs to be considered when mapping instruction forms to ports. This method is
currently used to validate known information, but can be extended to derive a complete, previously
unknown, port model.

\subsection{Example: Fused Multiply-Add on Skylake and Zen}\label{ssec:methodology_example}
To illustrate our model construction method, we carry out the analysis of the instruction form
\texttt{vfmadd132pd m128,xmm2,xmm1} (i.e., multiplying a packed double-precision value from
memory and \texttt{xmm1}, adding this to \texttt{xmm2} and storing the result in \texttt{xmm1}) for
the latest Intel and AMD architectures.

We use the port model for Skylake, shown in Figure~\ref{fig:skl_diagram}, and
Zen, as presented in Figure~\ref{fig:zen_diagram}. The benchmark files for
latency and throughput are generated automatically as shown in the previous
section. E.g., the  basic repetitive instruction form for the latency measurement is
\texttt{vfmadd132pd (\%rax), \%xmm0, \%xmm0}. 
All these instruction forms must be executed separately due to the read-after-write hazard between the
current target register and the future source register \text{xmm0}. The throughput
benchmark is generated analogously with independent registers as operands. 

Based on these files, we configure and run benchmarks for various levels of parallelism.
On AMD~Zen the output will look like this (note that we are using Intel operand ordering here
since ibench works with Intel assembly syntax internally):
\begin{lstlisting}[language={C}, keywordstyle=\color{red!70!black}\bfseries, commentstyle=\color{blue}, frame=lines, numbers=left, stepnumber=2, firstnumber=1, xleftmargin=2em]
Using frequency 1.80GHz.
vfmadd132pd-xmm_xmm_mem-1:         5.011 (clk cy)
vfmadd132pd-xmm_xmm_mem-2:         2.506 (clk cy)
vfmadd132pd-xmm_xmm_mem-4:         1.251 (clk cy)
vfmadd132pd-xmm_xmm_mem-5:         1.003 (clk cy)
vfmadd132pd-xmm_xmm_mem-8:         0.679 (clk cy)
vfmadd132pd-xmm_xmm_mem-10:        0.503 (clk cy)
vfmadd132pd-xmm_xmm_mem-12:        0.502 (clk cy)
vfmadd132pd-xmm_xmm_mem-TP:        0.500 (clk cy)
vfmadd132pd-xmm_xmm_xmm-TP:        0.502 (clk cy)
\end{lstlisting}
The number behind every instruction form is the amount of independent parallel instructions
in one loop iteration given the dependencies in every benchmark. ``TP'' marks
throughput benchmarks, without dependencies. On line~2, we can see that the latency of this
instruction form is 5\,cy. The reciprocal throughput shown on line~9 is 0.5\,cy/instr. The measured
throughput is unaffected for benchmarks with ten or more independent instruction forms, which
corroborates our general assumptions about multi-port code execution:
The instruction form can be spread among two separate ports,
because its throughput is one half and we expect each port to handle one instruction per cycle. Given
that Zen can do two loads per cycle and the instruction form without a memory operand has
a reciprocal throughput of 0.5\,cy/instr. as well,
we need to find which floating point ports (0, 1, 2 or 3) are
needed for fused-multiply-add (FMA). We therefore create benchmarks with
\texttt{vmulpd \%xmm1,\%xmm2,\%xmm3} and \texttt{vaddpd \%xmm1,\%xmm2,\%xmm3} instruction forms interleaved
with the prior \texttt{vfmadd132pd}. At this point, we already know that \texttt{vmulpd} is
executed on floating point port 0 or 1, \texttt{vaddpd} goes to port 2 or 3 and both instruction forms
have a reciprocal throughput of 0.5\,cy/instr. The chosen operands must be independent of the target
register to prevent hazards and therefore affect dependencies. The result is the following:
\begin{lstlisting}[language={C}, keywordstyle=\color{red!70!black}\bfseries, commentstyle=\color{blue}, frame=lines]
Using frequency 1.80GHz.
vfmadd132pd-xmm_xmm_mem-TP-vaddpd:    0.522 (clk cy)
vfmadd132pd-xmm_xmm_mem-TP-vmulpd:    1.024 (clk cy)
\end{lstlisting}
From the combined measurement we see that \texttt{vmulpd} -- unlike \texttt{vaddpd} -- can not be
hidden behind the execution of \texttt{vfmadd132pd}, so \texttt{vfmadd132pd} must be scheduled to
the same ports as \texttt{vmulpd}, i.e., 0 or 1. To add the instruction form to the Zen port model
of OSACA, we  create a new entry with a reciprocal throughput of 0.5\,cy/instr. on port 0, 1, 8 and 9 to
the database:
\begin{lstlisting}[language={C}, keywordstyle=\color{red!70!black}\bfseries, commentstyle=\color{blue}, frame=lines]
vfmadd132pd-xmm_xmm_mem, 0.5, 5.0, \
		  "(0.5,0.5,0,0,0,0,0,0,0,0.5,0.5)"
\end{lstlisting}
Note that for floating point division we assume that there is an additional divider pipe on port 3, which is included in the port occupation notation of the database.
For doing the same workflow on Skylake, we can reuse all priorly created benchmark codes and get
the following results:
\begin{lstlisting}[language={C}, keywordstyle=\color{red!70!black}\bfseries, commentstyle=\color{blue}, frame=lines, numbers=left, stepnumber=2, firstnumber=1, xleftmargin=2em]
Using frequency 1.80GHz.
vfmadd132pd-xmm_xmm_mem-1:         4.009 (clk cy)
vfmadd132pd-xmm_xmm_mem-2:         2.006 (clk cy)
vfmadd132pd-xmm_xmm_mem-4:         1.011 (clk cy)
vfmadd132pd-xmm_xmm_mem-5:         0.805 (clk cy)
vfmadd132pd-xmm_xmm_mem-8:         0.556 (clk cy)
vfmadd132pd-xmm_xmm_mem-10:        0.554 (clk cy)
vfmadd132pd-xmm_xmm_mem-12:        0.551 (clk cy)
vfmadd132pd-xmm_xmm_mem-TP:        0.553 (clk cy)
vfmadd132pd-xmm_xmm_xmm-TP:        0.502 (clk cy)
vfmadd132pd-xmm_xmm_mem-TP-vaddpd: 1.010 (clk cy)
vfmadd132pd-xmm_xmm_mem-TP-vmulpd: 1.004 (clk cy)
\end{lstlisting}
Here we get the same expected amount of cycles for the instruction form in combination with
\texttt{vaddpd} and \texttt{vmulpd} because both functional units are assigned to port 0 and 1,
increasing the overall throughput to 1\,cy. This leads to the assumption of a latency of 4\,cy
and a reciprocal throughput of 0.5\,cy/instr. The port distribution is 0, 1 for FMA, and 2, 3 for Load.
This is represented in the database in the following way:
\begin{lstlisting}[language={C}, keywordstyle=\color{red!70!black}\bfseries, commentstyle=\color{blue}, frame=lines]
vfmadd132pd-xmm_xmm_mem, 0.5, 4.0, \
		  "(0.5,0,0.5,0.5,0.5,0,0,0,0)"
\end{lstlisting}
Note that -- similar to Zen -- the Skylake architecture has an additional divider pipe on port 0.

Doing this for every instruction will give a validated port model, which follows the general
structure as seen in Figure~\ref{fig:skl_diagram} and Figure~\ref{fig:zen_diagram}.

\section{Static Analyzer Implementation}\label{sec:implementation}

After collecting the performance and scheduling information about specific instruction forms for a
given architecture, as done in Section~\ref{sec:methodology}, OSACA can use it to predict the
throughput of kernels.

OSACA extracts a marked kernel section out of an assembly or object file. For convenience OSACA
supports the same byte markers as IACA, i.e.:
\begin{lstlisting}[language={[x86masm]Assembler}, keywordstyle=\color{black}, commentstyle=\color{blue}, escapechar={ß}, frame=lines]
 ß\highl{red!70!black}{movl     \$111, \%ebx}ß
 ß\highl{red!70!black}{.byte    100,103,144}ß
     # ..LABEL:
	 # Some code
	 # ...
	 # conditional jump to ..LABEL
 ß\highl{red!70!black}{movl     \$222, \%ebx}ß
 ß\highl{red!70!black}{.byte	100,103,144}ß
\end{lstlisting}
These markers can be inserted in the source code, but we have found that this strongly
influences the code generated by the compiler. We therefore recommend to insert the marker instructions
directly into assembly code, to guarantee preservation of the original instructions.

\begin{figure*}[tbp]
\centering
\includegraphics[width=\textwidth]{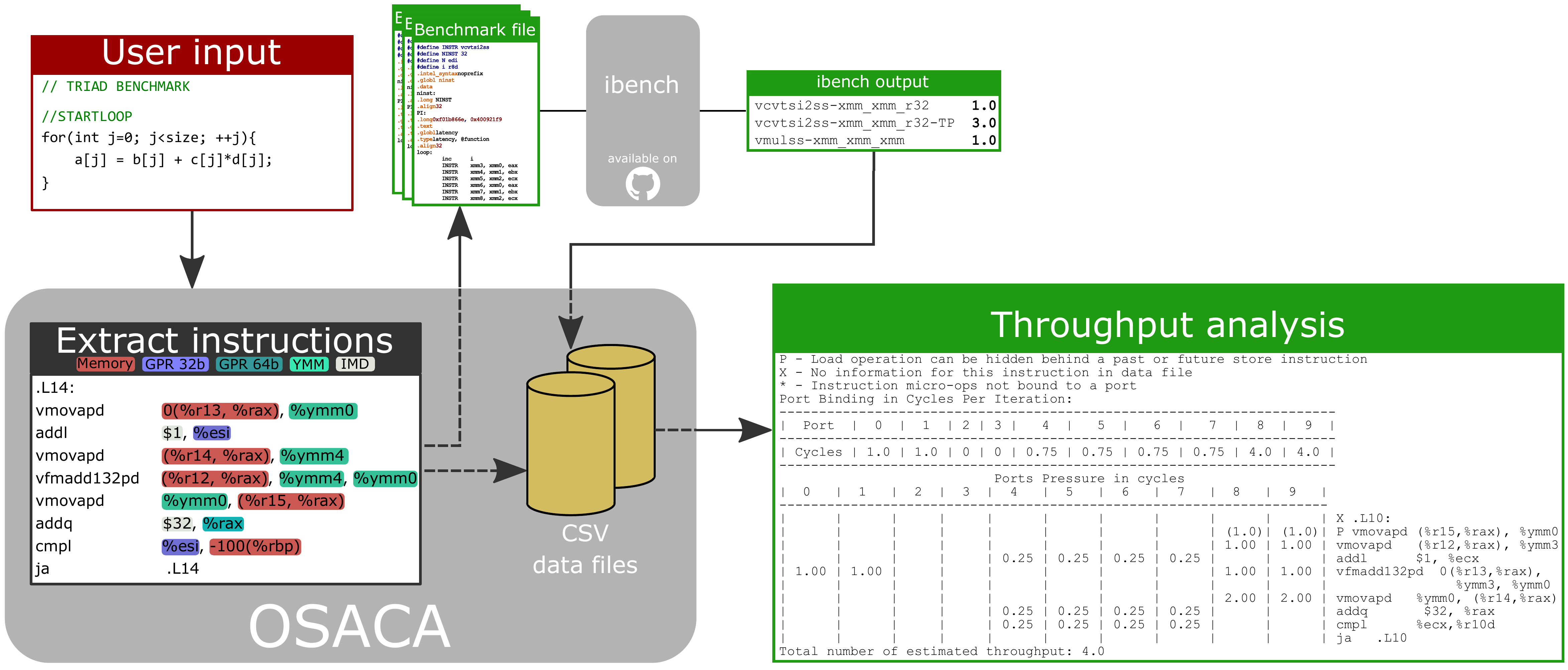}
\caption{Structural design of OSACA and its workflow. The code example shown here
  is the Sch\"onauer triad benchmark compiled with \texttt{-O3} and analyzed with
  OSACA assuming the Zen architecture.}
\label{fig:osaca}
\end{figure*}

Extracting the inner kernel is done using regular expressions. IACA's analysis is based on compiled
binary object files, which is an unnecessary step with OSACA. Each instruction form is analyzed
regarding its operands and matched to entries in the database. If no match was found, corresponding
benchmark files, as described in Section~\ref{ssec:latTp}, are generated automatically. If every
instruction form was found, OSACA performs a throughput analysis based on earlier measured data and
the port distribution from its database. The workflow of OSACA is depicted in
Figure~\ref{fig:osaca}.
For validation we will
use different assembly
representations, which are generated by the GNU C Compiler with different optimization levels:
\texttt{-O1}, \texttt{-O2} or \texttt{-O3}. The predictions by OSACA are validated by
comparing predicted runtime to measured execution time on the systems described in
Section~\ref{ssec:validation_hardware}. In case of Skylake we also compare OSACA and IACA
predictions, which is impossible for Zen due to the proprietary nature of IACA.

\subsection{Example: Triad on Skylake and Zen}\label{ssec:implementation_example}

A typical benchmark for measuring data throughput in combination with floating-point operations is
the ``Sch\"onauer'' triad benchmark~\cite{schoenauer00}:
\begin{lstlisting}[language={C}, keywordstyle=\color{blue!70!black}\bfseries, commentstyle=\color{blue}, frame=lines, escapechar={ß}]
for(int j=0; j<size; ++j)
	a[j] = b[j] + c[j]*d[j];
\end{lstlisting}
First, we analyze the kernel compiled with Skylake-specific optimization flags on both
architectures. Later, we will do the same analysis on both architectures with code compiled for
Zen. The resulting maximum measured number of floating operations~(FLOP) per second, the maximum number
of high-level, i.e., source code loop iterations (it) per second, and the minimum number of cycles per
iteration are stated in columns 5--7 of
Table~\ref{tab:zen-triad-meas}.\footnote{The relation between FLOPs and
  iterations is trivial in this example; for more complicated codes it is often useful to think
  in terms of iterations instead of FLOPs, so we keep both metrics.}

The FLOP/s metric is calculated from the total runtime and total number of FLOPs:
$2\frac{\mathrm{FLOP}}{\mathrm{iteration}}\times
\mathrm{\texttt{size}} \times \mathrm{repetitions} / \mathrm{runtime}$. It/s is calculated from
$\mathrm{\texttt{size}} \times \mathrm{repetitions} / \mathrm{runtime}$. Finally, the number of
cycles (cy/it) is calculated by dividing the clock speed (cy/s) by the
performance (it/s).

\begin{table}[tbp]
\centering
\begin{tabular}{c c c c c c}
\hline
Compiled for & Flag & unroll & \multicolumn{2}{c}{OSACA pred. [cy]} & IACA pred. [cy] \\
&& factor & Zen & SKL & SKL\\
\hline
\hline
Skylake & \texttt{-O1} & 1 & 2.00 & 2.00 & 2.24 \\
Skylake & \texttt{-O2} & 1 & 2.00 & 2.00 & 2.00 \\
Skylake & \texttt{-O3} & 4 & 4.00 & 2.00 & 2.21 \\
Zen 	& \texttt{-O1} & 1 & 2.00 & 2.00 & 2.24 \\
Zen 	& \texttt{-O2} & 1 & 2.00 & 2.00 & 2.00 \\
Zen 	& \texttt{-O3} & 2 & 2.00 & 2.00 & 2.21 \\
\hline
\end{tabular}
\caption{OSACA and IACA throughput analyses for the Sch\"onauer triad kernel. Note that the cycle
  counts pertain to one assembly loop iteration, which may comprise several source code
  iterations (according to the unroll factor).}
\label{tab:TPpreds}
\end{table}
The compiler unrolls the kernel four times at \texttt{-O3} for AVX SIMD vectorization (see
Figure~\ref{fig:osaca}).  Unrolling must be observed when interpreting IACA or OSACA
predictions, as they disregard the source code and only predict for assembly-level
iterations. E.g., if a loop was unrolled twice, the prediction by IACA and OSACA will be for two
original iterations instead of one. This also applies to any additional unrolling on top of SIMD.
In this paper, OSACA and IACA predictions given in cycles are for
one assembly code iteration, whereas the unit ``cy/it'' always refers to source
code iterations. 

All predictions by OSACA and IACA for ``Skylake-optimized'' code can be found in the first three
rows of Table~\ref{tab:TPpreds}.  OSACA's throughput analysis via \texttt{osaca --arch skl --iaca
  asmfile.s}, i.e., for Skylake, predicts 2\,cycles independent of the optimization level.
As mentioned above, OSACA predicts the throughput for one iteration of the marked kernel code,
which corresponds to one iteration in case of the \texttt{-O1} and \texttt{-O2} code and four
iterations in case of \texttt{-O3}. The OSACA prediction for the \texttt{-O3} code is shown
in somewhat condensed form in Table~\ref{tab:o3-skl-triad}.
\begin{table}[tbp]
\centering
{\scriptsize
\setlength\tabcolsep{2.5pt}
\begin{tabular}{cGcGcGcGl}
\hline
\rowcolor{white} P0 & P1 & P2 & P3 & P4 & P5 & P6 & P7 & Assembly Instructions\\
\hline
\hline
     &      &      &      &      &      &      &  & {\tiny \texttt{.L10:}}\\
     &      & 0.50 & 0.50 &      &      &      &  & {\tiny \texttt{vmovapd (\%r15,\%rax), \%ymm0}}\\
     &      & 0.50 & 0.50 &      &      &      &  & {\tiny \texttt{vmovapd (\%r12,\%rax), \%ymm3}}\\
0.25 & 0.25 &      &      &      & 0.25 & 0.25 &  & {\tiny \texttt{addl \$1, \%ecx}}\\
0.50 & 0.50 & 0.50 & 0.50 &      &      &      &  & {\tiny \texttt{vfmadd132pd 0(\%r13,\%rax),\%ymm3,\%ymm0}}\\
     &      & 0.50 & 0.50 & 1.00 &      &      &  & {\tiny \texttt{vmovapd \%ymm0, (\%r14,\%rax)}}\\
0.25 & 0.25 &      &      &      & 0.25 & 0.25 &  & {\tiny \texttt{addq \$32, \%rax}}\\
0.25 & 0.25 &      &      &      & 0.25 & 0.25 &  & {\tiny \texttt{cmpl \%ecx, \%r10d}}\\
     &      &      &      &      &      &      &  & {\tiny \texttt{ja .L10}}\\
\hline
\rowcolor{white} 1.25 & 1.25 & \textbf{2.00} & \textbf{2.00} & 1.00 & 0.75 & 0.75 & 0.00 \\
\hline
\end{tabular}}
\caption{OSACA prediction (shortened)
  of \texttt{-03} Sch\"onauer triad benchmark for Skylake with code compiled for
  Skylake. See Section~\ref{ssec:validation_hardware} for system configuration.}
\label{tab:o3-skl-triad}
\end{table}
Our measurement for the \texttt{-O3} code is $0.53$\,cy/it (see last row of
Table~\ref{tab:zen-triad-meas}), which matches both the OSACA and IACA predictions well since
\(4\,\mbox{it} \cdot 0.53\,\mbox{cy/it}= 2.12\,\mathrm{cy}\).

Unlike OSACA, IACA does not schedule instruction forms with an average probability but weighs
specific ports. The reason for this is not disclosed and may be based on internal information.
However, this does not affect the overall throughput and bottleneck prediction for the triad benchmark.
For the benchmark
versions compiled with \texttt{-O1}, \texttt{-O2} and \texttt{-O3}, IACA predicts between 2.00\,cy/it and 2.24\,cy/it for each kernel, but all with a pure port binding of 2.0\,cy in the bottleneck.
Running this code on Zen results in the same runtime as on Skylake for the \texttt{-O1}
and \texttt{-O2} versions, but shows worse performance with \texttt{-O3} (see rows
7--9 in Table~\ref{tab:zen-triad-meas}). OSACA's throughput
prediction for this version can be found in the structural design of Figure~\ref{fig:osaca}. The
lower performance is due to the Zen architecture executing AVX instructions as two successive
128-bit chunks.
This leads to an expected total runtime of 4 cycles per (assembly) iteration instead of Skylake's 2
(i.e., 1\,cy/it instead of 0.5), which is confirmed by the measurements in
column 7 of Table~\ref{tab:zen-triad-meas}.

The performance results for the triad benchmark compiled for the Zen architecture are shown in
the first six rows of Table~\ref{tab:zen-triad-meas}.
\begin{table*}[tbp]
\centering
\setlength\tabcolsep{5pt}
\begin{tabular}{c c c c|r r r|r r}
\hline
\rowcolor{white} \multicolumn{2}{c}{Architecture} & Optimization & Unroll & \multicolumn{3}{c|}{Measured} & \multicolumn{2}{c}{Prediction [cy/it]} \\
\rowcolor{white} executed on & compiled for & flag & factor & MFLOP/s & Mit/s & cy/it & OSACA & IACA \\
\hline
\hline
Zen     & Zen     &\texttt{-O1} & 1x & 1797 &  898 & 2.00 & 2.00 & --  \\
Zen     & Zen     &\texttt{-O2} & 1x & 1797 &  898 & 2.00 & 2.00 & --  \\
Zen     & Zen     &\texttt{-O3} & 2x & 3531 & 1754 & 1.02 & 2.00/2 & --  \\
\hline
Skylake & Zen     &\texttt{-O1} & 1x & 1770 &  885 & 2.03 & 2.00 & 2.24 \\
Skylake & Zen     &\texttt{-O2} & 1x & 1768 &  884 & 2.04 & 2.00 & 2.00 \\
Skylake & Zen     &\texttt{-O3} & 2x & 3505 & 1753 & 1.03 & 2.00/2 & 2.21/2 \\
\hline
Zen     & Skylake &\texttt{-O1} & 1x & 1792 &  896 & 2.01 & 2.00 & --  \\
Zen     & Skylake &\texttt{-O2} & 1x & 1797 &  898 & 2.01 & 2.00 & --  \\
Zen     & Skylake &\texttt{-O3} & 4x & 3166 & 1589 & 1.01 & 4.00/4 & --  \\
\hline
Skylake & Skylake &\texttt{-O1} & 1x & 1767 &  884 & 2.04 & 2.00 & 2.24 \\
Skylake & Skylake &\texttt{-O2} & 1x & 1776 &  888 & 2.03 & 2.00 & 2.00 \\
Skylake & Skylake &\texttt{-O3} & 4x & 6808 & 2738 & 0.53 & 2.00/4 & 2.21/4 \\
\hline
\end{tabular}
\caption{Measurements of the Sch\"onauer triad benchmark compiled for Intel Skylake and AMD Zen together with the corresponding predictions by OSACA and Intel IACA.}
\label{tab:zen-triad-meas}
\end{table*}
While we can observe similar behavior for the \texttt{-O1} and \texttt{-O2} versions compared to
the previous example, the compiler only unrolls twice for the \texttt{-O3} version, i.e.,
it only uses 128-bit wide registers. For all
six versions of the benchmark OSACA predicts 2\,cy per assembly iteration, which
matches the measured performance.
Since both architectures have the same throughput limits for 128-bit wide
data movement we do not see a
performance difference between Zen and Skylake.

The OSACA output for the \texttt{-O3} version compiled for Zen can be found in
Table~\ref{tab:o3-zen-triad}.
\begin{table}[tbp]
\centering
{\scriptsize
\setlength\tabcolsep{1pt}
\begin{tabular}{cGcGcGcGcGl}
\hline
\rowcolor{white} P0 & P1 & P2 & P3 & P4 & P5 & P6 & P7 & P8 & P9 & Assembly Instructions\\
\hline
\hline
     &      &      &      &      &      &      &      &       &       & {\tiny \texttt{.L10:}}\\
0.25 & 0.25 & 0.25 & 0.25 &      &      &      &      & (0.5) & (0.5) & {\tiny \texttt{vmovaps 0(\%r13,\%rax),\%xmm0}}\\
0.25 & 0.25 & 0.25 & 0.25 &      &      &      &      & 0.50  & 0.50  & {\tiny \texttt{vmovaps \%r15,\%rax),\%xmm3}}\\
     &      &      &      & 0.25 & 0.25 & 0.25 & 0.25 &       &       & {\tiny \texttt{incl \%esi}}\\
0.50 & 0.50 &      &      &      &      &      &      & 0.50  & 0.50  & {\tiny \texttt{vfmadd132pd (\%r14,\%rax),\%xmm3,\%xmm0}}\\
0.25 & 0.25 & 0.25 & 0.25 &      &      &      &      & 1.00  & 1.00  & {\tiny \texttt{vmovaps \%xmm0,(\%r12,\%rax)}}\\
     &      &      &      & 0.25 & 0.25 & 0.25 & 0.25 &       &       & {\tiny \texttt{addq \$16,\%rax}}\\
     &      &      &      & 0.25 & 0.25 & 0.25 & 0.25 &       &       & {\tiny \texttt{cmpl \%esi, \%ebx}}\\
     &      &      &      &      &      &      &      &       &       & {\tiny \texttt{ja .L10}}\\
\hline
\rowcolor{white} 1.25 & 1.25 & 0.75 & 0.75 & 0.75 & 0.75 & 0.75 & 0.75 & \textbf{2.0} & \textbf{2.0} \\
\hline
\end{tabular}}
\caption{OSACA prediction of \texttt{-03} Sch\"onauer triad benchmark for Zen with code compiled for Zen. Parentheses indicate a hide-able load \muop.}
\label{tab:o3-zen-triad}
\end{table}
Although Zen has two load units and one store unit on ports 8 and 9, it has only two AGUs on the very
same ports, so it is only capable of executing either up to two loads or one load and one store per
cycle. OSACA models this by hiding one load instruction behind a given store instruction, as seen on the
second row in Table~\ref{tab:o3-zen-triad} (\texttt{vmovaps 0(\%r13,\%rax),\%xmm0}).


\subsection{Example: $\pi$ Benchmark on Skylake and Zen}\label{ssec:implementation_example_2}

The Sch\"onauer triad benchmark is bound by load throughput. In the following we will look at an
arithmetic instruction bound benchmark that calculates $\pi = \int_0^14/(1+x^2)\,\mbox{dx}$ by
simple rectangular integration:
\begin{lstlisting}[language={C}, keywordstyle=\color{blue!70!black}\bfseries, commentstyle=\color{blue}, frame=lines, escapechar={ß}]
int SLICES = 1000000000;
double sum = 0., delta_x = 1./SLICES;
for(int i=0; i<SLICES; ++i) {
   double x = (i+0.5)*delta_x;
   sum = sum + 4.0 / ( 1.0 + x * x);
}
double Pi = sum * delta_x;
\end{lstlisting}
As in the previous example, we compiled the benchmark code with \texttt{-O1}, \texttt{-O2} and
\texttt{-O3} for Intel Skylake and AMD Zen with the flags described in
Section~\ref{ssec:validation_hardware}, but we only analyze and run on the architecture
we compile for. In order to convince GCC to vectorize this code with
\texttt{-O3}, the flag \texttt{-funsafe-math-optimizations} was required.
In Table~\ref{tab:results-pi} we have compiled IACA and OSACA predictions, as well as the measured
reciprocal throughput.

\begin{table}[tbp]
\centering
\begin{tabular}{ccccc}
\hline
\rowcolor{white} Arch. & Opt. & IACA & OSACA & Measurement \\
\hline
\hline
Skylake & \texttt{-O1} & 3.91\,cy/it &  4.75\,cy/it & 9.02\,cy/it \\
Skylake & \texttt{-O2} & 4.00\,cy/it &  4.25\,cy/it & 4.00\,cy/it \\
Skylake & \texttt{-O3} & 2.00\,cy/it &  2.00\,cy/it & 2.06\,cy/it \\
Zen     & \texttt{-O1} &             &  4.00\,cy/it & 11.48\,cy/it \\
Zen     & \texttt{-O2} &             &  4.00\,cy/it & 4.96\,cy/it \\
Zen     & \texttt{-O3} &             &  2.00\,cy/it & 2.44\,cy/it \\
\hline
\end{tabular}
\caption{Predictions and measurements of $\pi$ benchmark on Skylake and Zen.}
\label{tab:results-pi}
\end{table}

Predictions for \texttt{-O1} failed to describe the measured runtime by more than a factor of two
on both Skylake and Zen. Manual inspection of the code confirmed the validity of the
predictions under the model assumptions.
To investigate the discrepancy we checked the \texttt{UOPS\_EXECUTED\_STALL\_CYCLES}
hardware event using \texttt{likwid-perfctr} on Intel Skylake, and found that 
almost 17 times as many stall cycles were counted with \texttt{-O1} compared to \texttt{-O2}.
Measuring the average stall duration (part of
likwid's \texttt{UOPS\_ISSUE} group) also yields 5.5\,cy, which is roughly the
discrepancy we measured. Looking into the code again, the relevant difference is that at
\texttt{-O1}, the value of \texttt{sum} is read from the stack, updated, and written back
in every iteration:
\begin{lstlisting}[language={[x86masm]Assembler}, keywordstyle=\color{black}, commentstyle=\color{blue}, frame=lines, escapechar={ß}]
.L2:
  vxorpd	%xmm0, %xmm0, %xmm0	     
  vcvtsi2sd	%eax, %xmm0, %xmm0   
  vaddsd	%xmm4, %xmm0, %xmm0	     
  vmulsd	%xmm3, %xmm0, %xmm0	     
  vmulsd	%xmm0, %xmm0, %xmm0	     
  vaddsd	%xmm2, %xmm0, %xmm0	     
  vdivsd	%xmm0, %xmm1, %xmm0	     
  vaddsd	ß\bfseries(\%rsp)ß, %xmm0, %xmm5	     
  vmovsd	%xmm5, ß\bfseries(\%rsp)ß		     
  addl	$1, %eax		     
  cmpl	$1000000000, %eax	     
  jne .L2                             
\end{lstlisting}
It is kept in a register with \texttt{-O2} and only written back
after the loop.\footnote{All assembly kernels and the corresponding IACA and OSACA analysis
  can be found in the artifacts repository~\cite{artif}.}
We therefore conclude that on Skylake the write-after-read dependency
invalidates the full throughput assumption because of problems with the out-of-order
scheduler or speculative execution. On AMD
Zen the \texttt{DYN\_TOKENS\_DISP\_STALL\_CYCLES\_RETIRE\_TOKEN\_\\STALL} hardware event
points into the same direction, increasing to 7$\times$ to that of \texttt{-O2}, and thus we suspect that
the same problem exists there as well.

For \texttt{-O2} and \texttt{-O3}, predictions and measurements match rather well, in particular on
Intel Skylake. The throughput analysis for the \(\pi\) benchmark compiled for Skylake with
\texttt{-O3} and predicted for execution on Skylake can be found in Table~\ref{tab:o3-skl-pi}.
The compiler unrolled the kernel eight times, so that OSACA reports the model for
eight iterations of the loop. Note that for a precise prediction it is necessary to take a realistic 
execution of division \muops\ into account. Both Skylake and Zen use a different pipeline
for their divisions. Therefore, the ``main'' port is allocated only during one cycle of the execution,
while the remaining cycles leave the port free for other instructions.
OSACA supports division pipelines and marks them in the output as ``DV''.

\begin{table}[tbp]
\centering
{\scriptsize
\setlength\tabcolsep{2.5pt}
\begin{tabular}{ccGcGcGcGl}
\hline
\rowcolor{white} P0 &-- DV & P1 & P2 & P3 & P4 & P5 & P6 & P7  & Assembly Instructions\\
\hline
\hline
     &      &      &     &     &     &      &      &     & {\tiny \texttt{X .L2:}}\\
     &      &      &     &     &     & 1.00 &      &     & {\tiny \texttt{vextracti128   \$0x1, \%ymm2, \%xmm1}}\\
1.00 &      &      &     &     &     & 1.00 &      &     & {\tiny \texttt{vcvtdq2pd  \%xmm2, \%ymm0}}\\
0.50 &      & 0.50 &     &     &     &      &      &     & {\tiny \texttt{vaddpd \%ymm7, \%ymm0, \%ymm0}}\\
0.25 &      & 0.25 &     &     &     & 0.25 & 0.25 &     & {\tiny \texttt{addl   \$1, \%eax}}\\
1.00 &      &      &     &     &     & 1.00 &      &     & {\tiny \texttt{vcvtdq2pd  \%xmm1, \%ymm1}}\\
0.50 &      & 0.50 &     &     &     &      &      &     & {\tiny \texttt{vaddpd \%ymm7, \%ymm1, \%ymm1}}\\
0.33 &      & 0.33 &     &     &     & 0.33 &      &     & {\tiny \texttt{vpaddd \%ymm8, \%ymm2, \%ymm2}}\\
0.50 &      & 0.50 &     &     &     &      &      &     & {\tiny \texttt{vmulpd \%ymm6, \%ymm0, \%ymm0}}\\
0.50 &      & 0.50 &     &     &     &      &      &     & {\tiny \texttt{vmulpd \%ymm6, \%ymm1, \%ymm1}}\\
0.50 &      & 0.50 &     &     &     &      &      &     & {\tiny \texttt{vfmadd132pd    \%ymm0, \%ymm5, \%ymm0}}\\
0.50 &      & 0.50 &     &     &     &      &      &     & {\tiny \texttt{vfmadd132pd    \%ymm1, \%ymm5, \%ymm1}}\\
1.00 & 8.00 &      &     &     &     &      &      &     & {\tiny \texttt{vdivpd \%ymm0, \%ymm4, \%ymm0}}\\
1.00 & 8.00 &      &     &     &     &      &      &     & {\tiny \texttt{vdivpd \%ymm1, \%ymm4, \%ymm1}}\\
0.50 &      & 0.50 &     &     &     &      &      &     & {\tiny \texttt{vaddpd \%ymm1, \%ymm0, \%ymm0}}\\
0.50 &      & 0.50 &     &     &     &      &      &     & {\tiny \texttt{vaddpd \%ymm0, \%ymm3, \%ymm3}}\\
0.25 &      & 0.25 &     &     &     & 0.25 & 0.25 &     & {\tiny \texttt{cmpl   \$125000000, \%eax}}\\
     &      &      &     &     &     &      &      &     & {\tiny \texttt{jne    .L2}}\\
\hline
\rowcolor{white} 8.83 & \textbf{16.0} & 4.83 & 0.00 & 0.00 & 0.00 & 3.83 & 0.50 & 0.00 &  \\
\hline
\end{tabular}}
\caption{OSACA prediction of \texttt{-O3} \(\pi\)--benchmark for Skylake, compiled for Skylake.}
\label{tab:o3-skl-pi}
\end{table}

The fact that OSACA models the instruction throughput in average port occupation, as mentioned
before in Section~\ref{ssec:background}, leads to a prediction of 4.25\,cy instead of 4\,cy for the
execution of the benchmark compiled with \texttt{-O2} on Skylake (see Tables~\ref{tab:results-pi}
and~\ref{tab:o2-skl-pi-osaca}). According to IACA, \muops\ for instructions such as
\texttt{vxorpd} or \texttt{cmp} are not bound to a port, indicating that they take
``shortcuts'' through the architecture, avoiding port contention. This knowledge
is (still) lacking in OSACA, which leads to the OSACA in-core throughput model
not being a strictly lower bound for the execution time in all cases. This error is
generally small, however. 


In all other \texttt{-O2} and \texttt{-O3} predictions the division pipeline
on port 0 for Skylake (and port 3 for Zen, respectively) is the main throughput bottleneck of the
code.
\begin{table}[tbp]
\centering
{\scriptsize
\setlength\tabcolsep{2.5pt}
\begin{tabular}{ccGcGcGcGl}
\hline
\rowcolor{white} P0 &-- DV & P1 & P2 & P3 & P4 & P5 & P6 & P7  & Assembly Instructions\\
\hline
\hline
     &      &      &     &     &     &      &      &     & {\tiny \texttt{X .L2:}}\\
0.25 &      & 0.25 &     &     &     & 0.25 & 0.25 &     & {\tiny \texttt{vxorpd \%xmm0, \%xmm0, \%xmm0}}\\
0.50 &      & 0.50 &     &     &     & 1.00 &      &     & {\tiny \texttt{vcvtsi2sd	\%eax, \%xmm0, \%xmm0}}\\
0.25 &      & 0.25 &     &     &     & 0.25 & 0.25 &     & {\tiny \texttt{addl	\$1, \%eax}}\\
0.50 &      & 0.50 &     &     &     &      &      &     & {\tiny \texttt{vaddsd	\%xmm5, \%xmm0, \%xmm0}}\\
0.50 &      & 0.50 &     &     &     &      &      &     & {\tiny \texttt{vmulsd	\%xmm3, \%xmm0, \%xmm0}}\\
0.50 &      & 0.50 &     &     &     &      &      &     & {\tiny \texttt{vfmadd132sd	\%xmm0, \%xmm4, \%xmm0}}\\
1.00 & 4.00 &      &     &     &     &      &      &     & {\tiny \texttt{vdivsd	\%xmm0, \%xmm2, \%xmm0}}\\
0.50 &      & 0.50 &     &     &     &      &      &     & {\tiny \texttt{vaddsd	\%xmm0, \%xmm1, \%xmm1}}\\
0.25 &      & 0.25 &     &     &     & 0.25 & 0.25 &     & {\tiny \texttt{cmpl	\$1000000000, \%eax}}\\
     &      &      &     &     &     &      &      &     & {\tiny \texttt{jne	.L2}}\\
\hline
\rowcolor{white} \textbf{4.25} & 4.00 & 3.25 & 0.00 & 0.00 & 0.00 & 1.75 & 0.75 & 0.00 &  \\
\hline
\end{tabular}}
\caption{OSACA prediction of \texttt{-O2} \(\pi\)--benchmark for Skylake, compiled for Skylake.}
\label{tab:o2-skl-pi-osaca}
\end{table}
With AMD Zen, the execution is about 20\% slower than the prediction.  Just as on Skylake, we can
observe the bottleneck in the division pipeline for both the \texttt{-O2} and \texttt{-O3} version.

\section{Conclusion}\label{sec:conclusion}

\subsection{Summary}

Using our Open-Source Architecture Code Analyzer (OSACA)
we have shown that a partially automatic machine model construction and fully automatic throughput
analysis of loop kernels based on benchmarking and known hardware features is possible and yields
accurate results. Benchmarks are necessary to build a port model and gather throughput and latency
numbers of specific instruction forms. This approach yields deep insight into the in-core
performance limitations of a core micro-architecture. We verified our model, performance data and
predictions on Intel Skylake and AMD Zen CPU architectures using two kernels that show different
bottlenecks, and compared it with measured runtimes as well as predictions from Intel's
Architecture Code Analyzer (IACA).

OSACA can extract loop kernels and analyze their instruction forms out of marked assembly code.
Using techniques shown in this work, one can refine the port models and create realistic best-case
throughput predictions for in-core execution. OSACA is intended as an alternative to IACA,
with the ability to go beyond Intel hardware. 

\subsection{Future Work}\label{ssec:future}

OSACA in its current state is a first draft of what we envision for the future. We intend to extend
it with various new core features, the most relevant one being latency modeling (which has been
dropped by IACA some years ago). This requires support for critical path analysis, tracking
dependencies between sources and destinations as well as a model for output
forwarding. Differentiation between memory addressing modes is already part of the design of OSACA,
but not completely implemented. This is crucial for modeling the peculiar AGU behind port 7 on
Haswell and beyond, and more generally for any architecture where different addressing modes
can have varying performance impact.

Once there is a solid model of all the disclosed and known features, we will also start including
less well understood behavior, such as heuristics of the out-of-order scheduler and ``shortcuts''
that bypass the port scheduler. This will involve
very detailed and precise benchmarking to identify the underlying rules. Furthermore, we
intend to add information about the critical path and loop carried dependencies to enhance the
lower bound model in a realistic way.

Support for non-x86 architectures is on the horizon but will require a lot of manual model
building and validation, since they are not as well understood from a performance perspective. In
order to support these efforts, we are pursuing an automatic deduction approach, which is in early
development.

A framework for easier and more flexible generation of benchmarks is currently in development.
\cite{asmjit_poster}.

\section*{} 

\bibliographystyle{IEEEtran}
\bibliography{IEEEabrv,citations}

\end{document}